\documentclass{aa}
\usepackage{psfig}
\usepackage{graphicx}
\usepackage{times}
\begin{document}

%\thesaurus{11     % Galaxies
%              (11.03.2;  % Galaxies: compact,
%               11.09.2;  % Galaxies: interactions,
%               11.05.2)}  % Galaxies: evolution.

\title{On faint companions in the close environment of star--forming dwarf galaxies}
\subtitle{Possible external star formation triggers ?}

\author{K.G. Noeske \inst{1}
\thanks{Summer research student 1998, Instituto de Astrof\'{\i}sica de Canarias}
        \and
        J. Iglesias-P\'{a}ramo\inst{2}
        \and
        J.M. V\'{\i}lchez\inst{3}
        \and
        P. Papaderos\inst{1}
        \and
        K.J. Fricke\inst{1}          
}

\offprints{K. Noeske, \email{knoeske@uni-sw.gwdg.de}}

\institute{
Universit\"{a}ts-Sternwarte G\"{o}ttingen, D-37083 G\"ottingen, Germany
\and
Instituto de Astrof\'{\i}sica de Canarias, 38200 La Laguna, S/C de Tenerife, Spain
\and
Instituto de Astrof\'{\i}sica de Andaluc\'{\i}a (CSIC), 18080 Granada, Spain
}

\date{Received 25 September 2000 / Accepted 12 March 2001}

%----------------------------------------------------------------------------------
\abstract{
We have searched for companion galaxies in the close environment of 98 
star--forming dwarf galaxies (SFDGs) from field and low density environments, 
using the NASA Extragalactic Database.
Most of the companions are dwarf galaxies which due to
observational selection effects were previously disregarded in 
environmental studies of SFDGs. 
A subsample at low redshift, $cz <$ 2000 \,km\,s$^{-1}$, was chosen  to 
partially eliminate the observational bias against distant dwarf companions.
We find companion candidates for approximately 30\,\% of the objects
within a projected linear separation $s_p <$100\,kpc and a redshift difference 
$\Delta cz <$ 500\,km\,s$^{-1}$.
The limited completeness of the available data sets, 
together with the non-negligible frequency of \ion{H}{i}
clouds in the vicinity of SFDGs indicated by recent radio surveys, suggest that a
considerably larger fraction of these galaxies may be accompanied by low--mass systems. 
This casts doubt on the hypothesis that the majority of them can be considered truly 
isolated.
The velocity differences between companion candidates and sample SFDGs amount 
typically to $\la$ 250 km\,s$^{-1}$, and show a rising distribution towards lower
$\Delta cz$. This is similarly found for dwarf satellites of spiral galaxies, 
suggesting a physical association between the companion candidates and the sample SFDGs.
SFDGs with a close companion do not show significant differences in their 
H$\beta$ equivalent widths and $B-V$ colours as compared to isolated ones. However,
the available data do not allow us to rule out that interactions with close dwarf 
companions can influence the star formation activity in SFDGs.
}
%________________________________________________________________
\maketitle
\authorrunning{K. Noeske et al.}
\titlerunning{Faint companions in the close environment of star--forming dwarf galaxies}
%
%----------------------------------------------------------------------------------
\section{Introduction}
\label{intro}
%----------------------------------------------------------------------------------
%
Dwarf galaxies are considered important contributors to the baryonic mass of the 
Universe and to the star formation rate (SFR) density at higher redshifts 
(see. e.g. Guzm\'an et al. \cite{guzman98}). Some of them may even be similar 
to low--mass building blocks of normal galaxies in a bottom--up cosmological model.
A better understanding of the processes that drive their evolution is therefore central
to cosmological studies, and to the understanding of galaxy formation and evolution.

Starbursts, i.e. brief episodes of strongly enhanced star formation (SF) activity, 
are thought to occur frequently during the lifetime of a gas--rich dwarf galaxy.
A number of internal processes has been put forward to explain their origin and 
transient nature, such as Stochastic Self--Propagating Star Formation 
(Gerola, Seiden \& Schulmann \cite{gerola80}, see Thuan \cite{thuan91})
or a cyclic process of gas infall onto and expulsion from an older stellar host
(e.g. Davies \& Phillips \cite{davies88}, Papaderos et al. \cite{papaderos96}).   

Alternatively, environmental influences have been suggested to induce
 starburst activity in star--forming dwarf galaxies (SFDGs; see Section \ref{data}).  
Based on the hypothesis that dwarf galaxies are affected in a similar way to  
luminous disk galaxies (e.g. Kennicutt et al. \cite{kennicutt87}), mainly tidal 
interactions with stellar or gaseous companions have been treated by former studies.

Information on the presence and frequency of such external perturbors has been assembled 
from environmental studies of emission line galaxies (ELGs), performed mainly in the 
context of large scale structure formation and biased formation and evolution of galaxies 
in different environments. ELGs were found to smoothly follow the structures delineated 
by  luminous galaxies, although 
they are more fuzzily distributed (Rosenberg et al. \cite{rosenberg94}), generally populating 
lower-density environments (Salzer \cite{salzer89}, Telles \& Terlevich 1995, hereafter 
\cite{telles95}), and are less strongly clustered than less active luminous galaxies 
(Loveday et al. \cite{loveday99}, Telles \& Maddox \cite{telles99}, Lee et al. \cite{lee00}).
As for SFDGs, they were found to 
be concentrated towards void boundaries (Lindner et al. 1996, hereafter \cite{lindner96}), 
with a fraction ($\approx$ 20\% ) of them inside voids 
(Pustilnik et al. \cite{pustilnik95}), 
arranged in loose groups  which do not contain any bright galaxies (\cite{lindner96}). 
Interactions with {\em luminous} companions were therefore considered too rare to 
be the generic triggering agent of starbursts in SFDGs, given that luminous galaxies  
are rare in the close neighbourhood of SFDGs.

Whether interactions with companion galaxies can significantly influence SF activity 
and evolution of SFDGs has been investigated in earlier work (Campos--Aguilar et al. 
\cite{campos91}, \cite{campos93}; \cite{telles95}). 
The spectrophotometric properties of the studied objects showed no significant
dependence on the presence or absence of a companion galaxy, and both SF activity and 
metallicity seemed unrelated to distance or absolute magnitude of putative 
companions.
Campos--Aguilar et al. chose their samples 
from the spectrophotometric catalogue of \ion{H}{ii} galaxies (SCHG, Terlevich et al. 
\cite{terlevich91}), which mostly covers redshifts $cz\ga$ 2000 kms$^{-1}$. 
Taking into account the magnitude cutoff of the CfA catalogue used for the companion 
search (13.5 mag), these studies were practically restricted to luminous companions 
(M$\la$ --18 mag), typically not particularly close to SFDGs. 
\cite{telles95}, on their part, explicitly focus on the possible influence of 
luminous companions (M$<$ --19 mag). 
Also, the opposite conclusion by Grogin \& Geller (\cite{grogin00}), 
that ELGs in low--density environments show enhanced SF activity in the presence 
of a companion, is mostly drawn from studies of luminous ELGs and companions.

Observational evidence of a different kind points towards a correspondence between 
the membership of a SFDG in a given environment and its spectrophotometric
properties. A comparative study of SF activity in SFDGs populating different 
environments by V\'{\i}lchez (\cite{vilchez95}, \cite{vilchez97}) suggests that 
objects in low density environments have higher SF activity than 
those located in high density regions. The results of Hashimoto et al. 
(\cite{hashimoto98}) point in the same direction.
Popescu, Hopp and Rosa (\cite{popescu99}) reported no conspicuous difference in 
present and past SF activity for ELGs in field and void environments. On the other hand, 
Vennik, Hopp and Popescu (\cite{vennik00}) found {that the low surface brightness 
hosts of non--isolated ELGs tend to be more compact than those of isolated ones.

The mixed evidence for an influence of the environment on the 
SF activity of a SFDG calls for further studies, 
extending previous ones to larger samples and fainter magnitudes. 
This seems particularly important in view of the conjecture by
\cite{lindner96}, who attributed the apparent isolation of a fraction of BCDs 
to an observational bias against their low--luminosity companions.

Assuming that companions are capable of inducing SF activity in SFDGs,
 the mechanism and the prime parameters controlling this process remain unclear.
Tidal forces are, however, considered to be of major importance.
If $D_{c}$ and $M_{c}$ denote respectively the distance and mass of a companion, 
the tidal forces acting on a SFDG scale as
\begin{equation}
F_{tid}\propto M_{c}\times D_{c}^{-3},
\end{equation}
so that a nearby low--mass, i.e. faint, perturbor
can affect a SFDG to the degree a more distant giant 
galaxy does (cf. Campos--Aguilar et al. \cite{campos91}). 
Observational data (Wilcots et al. \cite{wilcots96}) 
and numerical models (Hensler et al. \cite{hensler99}, Pilyugin \cite{pilyugin00}) 
further suggest that, in addition to tidal interactions, low-mass stellar or gaseous 
companions may trigger and fuel starbursts by infalling onto a SFDG.

Although Telles \& Maddox (\cite{telles99}), from an analysis of APM catalogues, 
found no excess of dwarf companions for \ion{H}{ii} galaxies down to 
$\approx$ --14.5\,B\,mag and $\sim$ 10$^8$ M$_{\sun}$, numerous results point to the
presence of objects of even lower masses and luminosities.
Non--catalogued objects of low optical luminosities and surface brightnesses were
found for most one--armed magellanic irregulars (Odewahn  \cite{odewahn94}), and 
around many BCDs (Pustilnik et al. \cite{pustilnik97}, Walter et al. \cite{walter97}, 
Doublier et al. \cite{doublier99}, M\'endez \& Esteban 1999). 

Likewise, \ion{H}{i} observations of the environment of magellanic irregulars
(Wilcots et al. \cite{wilcots96}) and \ion{H}{ii} galaxies (Taylor et al. 
\cite{taylor94}, \cite{taylor95}, \cite{taylor96}) revealed
\ion{H}{i} companions for the majority of them, down to intergalactic \ion{H}{i} 
clouds with masses of $\sim$ 10$^7$ M$_{\sun}$ for some of which no optical 
counterpart was found.

The presence of still undetected objects of low optical and \ion{H}{i} 
luminosity appears also likely, keeping in mind that the LF steepens towards the 
low--luminosity end for SFDGs (Loveday et al. \cite{loveday99}), and in view of e.g. 
the high space density and the masses of local Ly$\alpha$ absorbers 
(Shull et al. \cite{shull96}).
Grogin \& Geller (\cite{grogin00}) found for luminous ELGs that
the frequency of companions at low redshift differences %{\em (so in GG) } 
is widely independent of the local galaxy density. Therefore, faint 
companions may not be rare even in regions of low density of luminous galaxies.

The aim of this study is to investigate the incidence of close 
optical companions of SFDGs and whether such close companions strongly affect
a SFDG's SF activity. Particular attention is attached to faint close companions, 
the presence and influence of which was generally not assessed in previous work. 
In order to address the problem of 
observational bias against low--luminosity sources, we investigate separately a subset of 
nearby SFDGs with recession velocities $<$ 2000 kms$^{-1}$. 
For a sample of SFDGs, we compiled photometric and spectroscopic properties 
at different wavelengths and corrected distance dependent observables 
for the Virgocentric infall.

The paper is organized as follows: In Section \ref{data} we describe our galaxy 
sample and the data processing. In Section \ref{results}, we analyse our data sets by 
applying different statistical methods and list the results, which are further 
discussed in  Section \ref{discussion}. In Section \ref{summary}, we summarize our work
and conclusions.
We assume $H_0 =$75\,km\,s$^{-1}$\,Mpc$^{-1}$ throughout.
%
%----------------------------------------------------------------------------------
\section{The Data}
\label{data}
%----------------------------------------------------------------------------------
\subsection{Sample selection}
\label{sample}

Various criteria were applied to select samples of dwarf galaxies with current 
or recent strong SF, which resulted in catalogues of Blue Compact Dwarf Galaxies 
(BCDs, e.g. Thuan \& Martin \cite{thuan81}) and \ion{H}{ii} galaxies 
(e.g. SCHG, Terlevich et al. \cite{terlevich91}).
Despite the different classification schemes,
it has been subsequently shown that the 
underlying stellar hosts of starbursting dwarf galaxies have generally a compact structure 
(Papaderos et al. \cite{papaderos96}, Marlowe et al. 1997, Salzer \& Norton \cite{salzer98}).
Also comparative studies of dwarf \ion{H}{ii} galaxies, BCDs and
dwarf amorphous galaxies with ongoing SF showed them to widely share 
the colours and structural properties of their host galaxies, relative starburst
luminosities and $EW$(H$\alpha$), suggesting that they basically form one and the same 
class of extragalactic objects (Papaderos et al. \cite{papaderos96}, Telles et al. 1997,
Marlowe et al. \cite{marlowe99}). 
We will therefore unify both BCDs and \ion{H}{ii} galaxies using
the term ``star--forming dwarf galaxies'' (SFDGs), 
following V\'{\i}lchez (\cite{vilchez95}, \cite{vilchez97}). 

In order to study a sufficiently large SFDG sample for which comparable 
observational data are available, we decided to use lists IV and V of
the sample of ELGs from the University of Michigan (UM) Survey (McAlpine et al. \cite{mcalpine77a},
 \cite{mcalpine77b}, \cite{mcalpine77c}; McAlpine \& Lewis \cite{mcalpine78};
McAlpine \& Williams \cite{mcalpine81}), systematically investigated in Salzer 
et al. (\cite{salzer89a}, \cite{salzer89b}) by means of spectroscopy
and CCD imaging in $B$ and $V$.
Among the different types of ELGs, all objects classified as \ion{H}{ii} galaxies 
were selected; the magnitude limit of  M$_B >$\,--18\,mag, commonly adopted to 
select dwarf galaxies, was applied {\em after} the distance correction 
({Section \ref{vc_correction}) had been performed.
More than 80\% of these UM galaxies are located at redshifts $cz>$\,2000\,km\,s$^{-1}$. 

Our sample  further includes SFDGs studied by Cair\'os
and Noeske (cf. Cair\'os \& V\'{\i}lchez \cite{cairos98}, Cair\'os et al. 
\cite{cairos00}, Noeske et al. \cite{noeske98}, \cite{noeske00a}, Noeske \cite{noeske99}). 
These are, for the most part, nearby objects
($cz <$ 2000\,km\,s$^{-1}$) which allow an environmental search that is less biased 
against faint objects than the quite distant UM sample.
None of our sample galaxies are members of galaxy clusters. Those which 
 are known or likely members of groups  
(cf. NED and Garcia \cite{garcia93}) were rejected (Mkn\,35, Mkn\,71 and Mkn\,527). 
Only 3 galaxies, UM\,454, UM\,455 and UM\,513, have been reported to lie
inside voids (Salzer \cite{salzer89}).
The resulting sample consists of 98 dwarf galaxies in typical field-- and low density 
environments and is listed in Table~\ref{samtab}.

%----------------------------------------------------------------------------------
\subsection{Search for companion objects}
\label{search}

Our companion search catalogue was the NASA Extragalactic Database
\footnote{The NASA/IPAC Extragalactic Database (NED) is operated by the Jet 
Propulsion Laboratory, California Institute of Technology, under contract 
with the National Aeronautics and Space Administration.} (NED). 

Most of our objects have declinations $\delta >$ --3$\degr$30$\arcmin$,
i.e. lie within the sky region covered by the
CGCG (Zwicky \cite{zwicky61}), which is included into the NED (cf. \cite{telles95}). 
We can therefore attribute the completeness limit of the CGCG, m$_B^{lim} 
\approx$ m$_{phot}^{lim} \approx $\,15.5\,mag, to our companion search.
As the NED includes numerous other literature sources, considerably fainter 
objects should partly be detectable, which makes NED a suitable catalogue to find 
as many known faint sources as possible. 
The non--uniform completeness of this database does, on the
other hand, not constitute a drawback since a meaningful extrapolation beyond even a 
well--defined completeness limit is at present not possible (see Section \ref{compprops}).

In an initial selection procedure, we considered any type of extragalactic
source a possible companion. Statistical analyses of spatial companion frequencies 
from excess source densities around SFDGs have been previously done
(Telles \& Maddox \cite{telles99}).
We hence restricted ourselves to sources for which redshift data was available. Thereby some
 companions are missed, but information on individual companions' properties such as
luminosity etc. can be analyzed.

Any object within a projected separation $s_p \leq 0.1$\,Mpc 
from our sample galaxies was included into a tentative list of companions. 
The corresponding angular search radius was calculated 
from the redshift $cz$ given in the NED, assuming a pure Hubble flow. We allowed, in 
the first place, a maximum separation in velocity space of 
$\Delta cz \leq 2000$\,km\,s$^{-1}$ between sample object and companion. 
The final selection criterion below which  
we finally considered an object a companion was, however, 
$\Delta cz \leq 500$\,km\,s$^{-1}$ (cf. Section \ref{distribution}).

%----------------------------------------------------------------------------------
\subsection{Distances and correction for Virgo Cluster infall}
\label{vc_correction}

The choice of a less distant subsample of SFDGs with $cz <$\,2000\,km\,s$^{-1}$ 
does not justify the assumption of a pure Hubble flow throughout to determine galaxy distances, 
but requires a correction for the peculiar velocity field in the environment of the 
Virgo Cluster (VC). For this purpose, we adopted the distances listed in the Nearby Galaxies 
Catalog (Tully 1988, hereafter \cite{tully88}). These are based on a Virgocentric infall model, 
which uses redshift -- independent distance determinations, morphology and neighbourhood 
considerations to treat those galaxies affected by the triple value problem 
(see Tully \& Shaya \cite{tully84}).

For all objects not listed in \cite{tully88}, the distance was directly calculated from their 
redshift given in the NED. Galaxies within the redshift range covered by \cite{tully88}, 
$cz <$\,3000\,km\,s$^{-1}$, which are not listed in the catalogue, were found to be at least  
1.5 mag fainter than the dwarf -- non dwarf separation limit of M$_B >$\,--18\,mag, i.e. 
can be considered dwarfs despite their possibly larger distance insecurities.
For galaxies outside the redshift range of \cite{tully88}, the deviation of the true distance 
from  the one obtained assuming a pure Hubble Flow can be estimated from Kraan--Korteweg 
(\cite{kraan86}, cf. their Fig. 3b) to $\la$\,30\% due to VC infall effects. 
A comparable deviation was obtained by
Marinoni et al. (\cite{marinoni98}, cf. their Fig. 14) who have adopted in their
study a multiattractor model. The above mentioned dispersion translates in the
worst case into an  uncertainty of $\la $ 0.6 mag in the distance modulus which may 
then slightly affect the luminosity--based  distinction between dwarf and
luminous galaxies. This does not affect the distance-independent 
quantities $B-V$ and $EW($H$\beta )$ involved in our study.

For the companion candidates, we assumed the same distances as for the sample galaxies 
around which they were found. 
Given that the peculiar velocity field in the VC vicinity
is not expected to change dramatically on scales of few tens of kpc, and that 
the companions show signs of physical association with the sample galaxies 
(cf. Section \ref{distribution}), the above assumption appears justified.
For these distances, projected separations $s_p$ between sample 
objects and candidate companions were calculated from their angular separation.

%----------------------------------------------------------------------------------
\subsection{Photometric and spectroscopic data}
\label{photometry}

Aperture photometry in optical broadbands was taken from the NED 
i.e. listing mostly RC3 data, from Salzer (\cite{salzer89a}, \cite{salzer89b}), 
Cair\'os et al. (\cite{cairos00}), Noeske (1999), 
and Noeske et al. (\cite{noeske98}, \cite{noeske00a}).
H$\beta$ equivalent widths were compiled from French (\cite{french80}), Thuan \& Martin 
(\cite{thuan81}), the SCHG (Terlevich et al. \cite{terlevich91}) and Salzer 
(\cite{salzer89a}, \cite{salzer89b}).

Adopting the distances obtained as described in Section \ref{vc_correction}, apparent 
magnitudes were transformed to absolute values.
A correction for galactic foreground dust extinction was applied to all photometric 
data using the $B$ band extinction given in the  NED for each galaxy, and  
applying the standard reddening curve by Savage \& Mathis (\cite{savage79}).

%----------------------------------------------------------------------------------
%----------------------------------------------------------------------------------
\section{Results}
\label{results}
%----------------------------------------------------------------------------------
%----------------------------------------------------------------------------------
\subsection{The close environment}

%-------------------------------------------
\begin{table}[!t]
\caption[]{Companion search results for subsamples within different redshift intervals} 
\label{biastab}
\begin{tabular}{lccccc}
\hline
$cz^a$ & sample & non&  no. of & dwarf \\
range & size & isol. & comps.&  comps. \\
(1)     & (2)   & (3)   & (4)   & (5) \\
\hline
unconstrained            & 98  & 16 & 18 & 15 \\
$cz < 2000$           & 42   & 13 & 15 & 14 \\
$2000 \leq cz < 4000$\hspace{0.5cm} & 12   & 2  & 2  & 1 \\
$4000 \leq cz < 6000$ & 20   & 0  & 0  & 0 \\
$6000 \leq cz < 8000$ & 15   & 1  & 1  & 0 \\
$cz \geq 8000$        & 9    & 0  & 0  & 0 \\
\hline \\
\end{tabular}
\\$^a$ redshifts are corrected for Virgo infall by multiplying 
the corrected distances from Section \ref{vc_correction} by $H_0$ (75\,km\,s$^{-1}$\,Mpc$^{-1}$).\\  
%$^b$ criterion for bound pair of galaxies\\
(1) redshift interval (in km\,s$^{-1}$) of respective subsample\\
(2) number of sample galaxies in subsample\\
(3) number of galaxies in the subsample for which at least one possible 
companion was found\\
(4) total number of possible companions found for the respective subsample\\
(5) number of dwarf galaxies among the possible companions\\
\end{table}
%--------------------------------------------

%%--------------------------------------------------------------------
\subsubsection{Distribution of the companions and companion selection criteria}
\label{distribution}

\begin{figure}
\resizebox{\hsize}{!}{\includegraphics{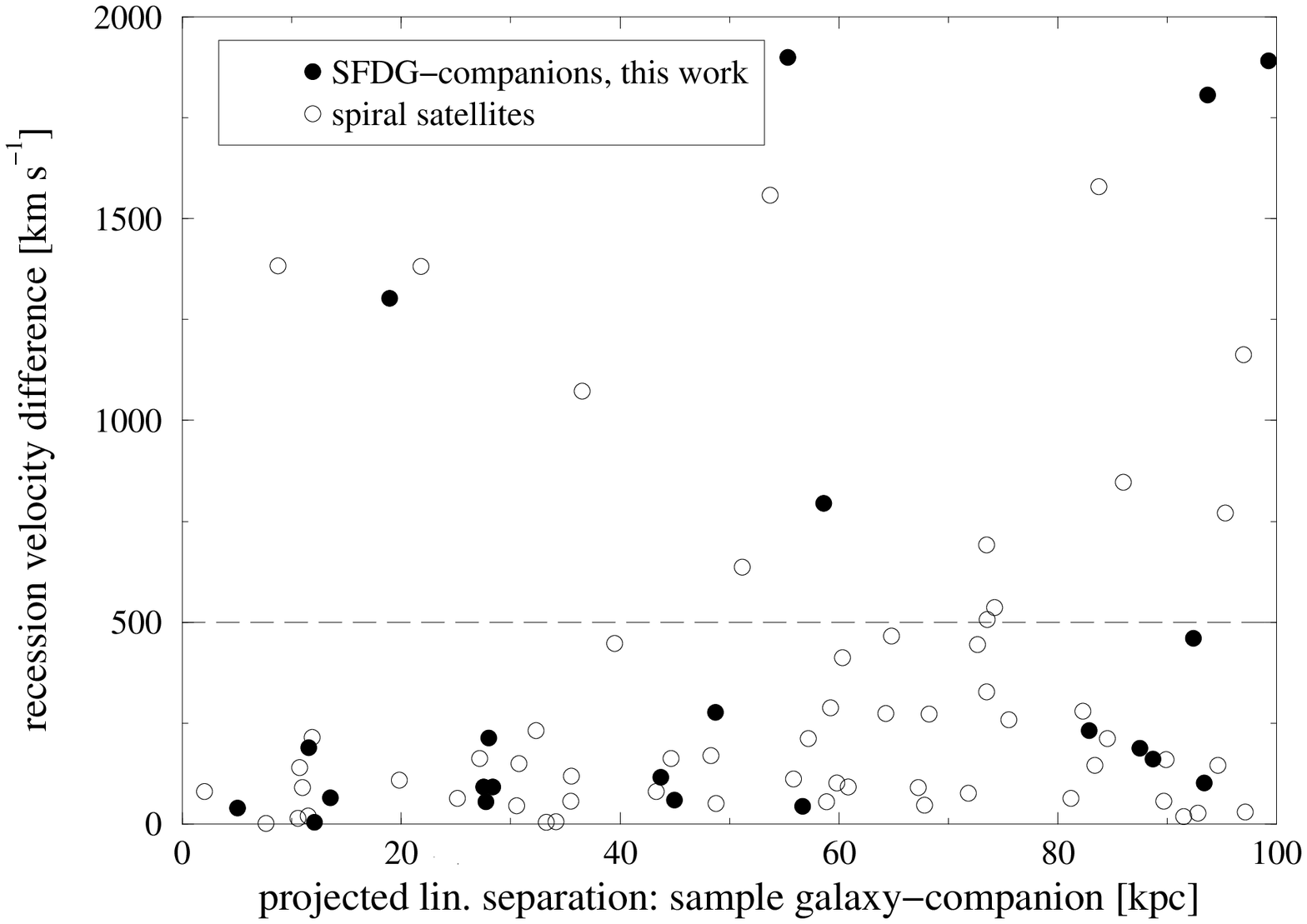}}
\caption{Recession velocity difference vs. projected linear 
separation of the SFDGs' putative companions (filled symbols). 
For comparison, we show by open circles the distribution of 
all companions (not restricted to dwarf galaxies) we found around 
field spiral galaxies from the sample of  Kennicutt \& Kent 
(\cite{kennicutt83}). The dashed line represents the maximum recession 
velocity difference between SFDG and companion we adopt.}
\label{tdv}
\end{figure}

\begin{figure}
\resizebox{\hsize}{!}{\includegraphics{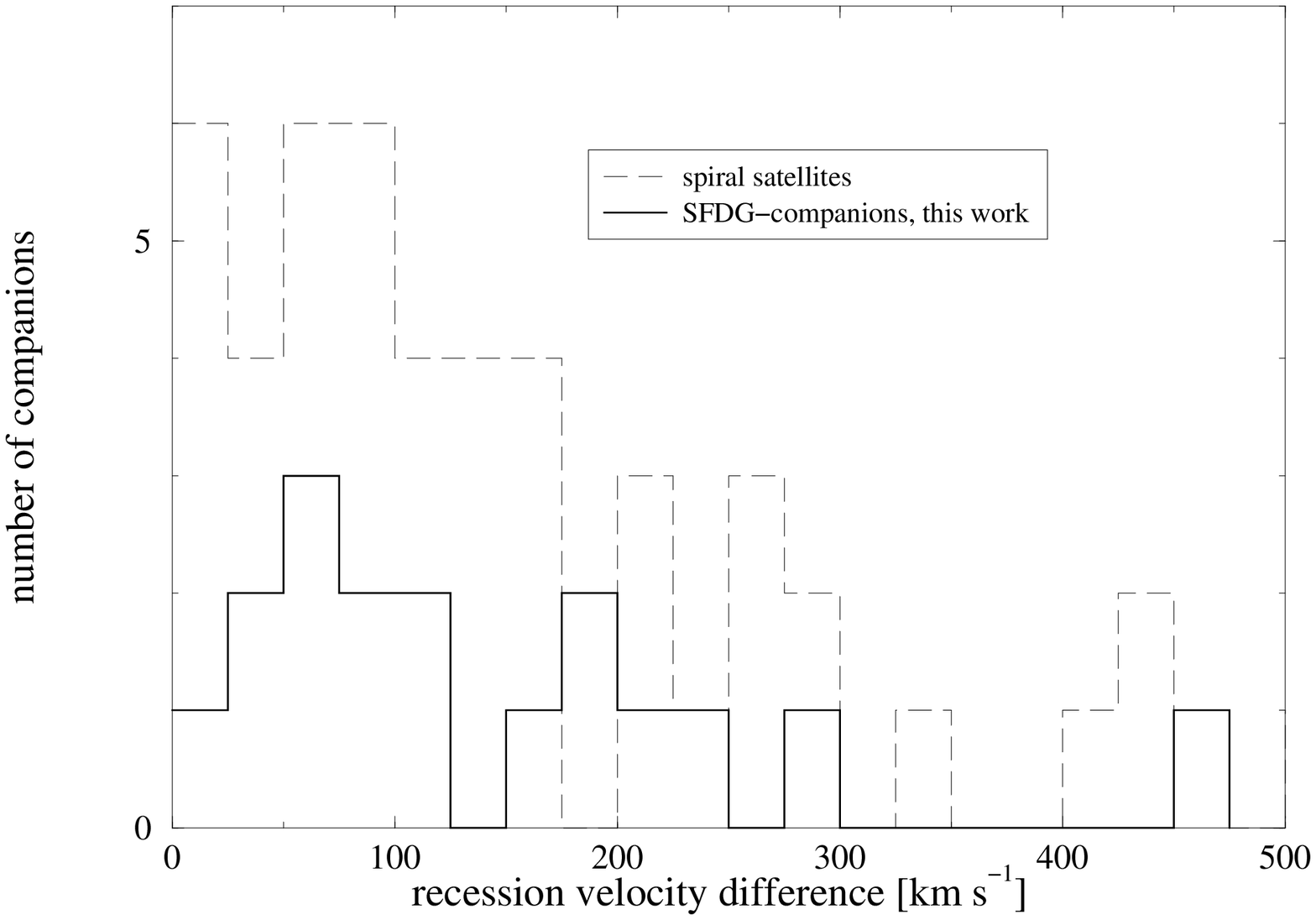}}
\caption{Distribution of the recession velocity differences between the sample SFDGs and 
the putative companions. The dashed histogram is built from the spiral companions shown in  
Figure \ref{tdv}.}
\label{com_dv}
\end{figure}

It is conceivable that the recession velocity difference, $\Delta cz$, 
and the projected separation, $s_p$, strongly influence the dynamical response and 
spectrophotometric evolution of dwarf galaxies in the presence of a companion.

We assumed tidal forces to have significant dynamical influence within the range 
$\Delta cz < $\,500\,km\,s$^{-1}$ and $s_p\,<$\,0.1\,Mpc.
These criteria follow the estimates by Campos--Aguilar et al. (\cite{campos93}), 
and are comparable to those by Pustilnik et al. (\cite{pustilnik00}).
They are further compatible with observed values of $s_p$ and $\Delta cz$  
for e.g. the BCD \object{II\,Zw\,33} and its companion \object{II\,Zw\,33\,B}, 
which are probably bound (Walter et al. \cite{walter97}) and likely to be influenced 
by interaction (M\'endez et al. \cite{mendez99}). 
For M51--like close pairs of galaxies, which show signs of perturbation due to
interaction, similar values for $\Delta cz$ are observed 
(Laurikainen et al. \cite{laurikainen98}).

The distribution of our SFDG companion candidates   
on the $\Delta cz$ vs. $s_p$ plane (Figure \ref{tdv}, filled circles) peaks  
towards small redshift differences. This is also illustrated 
in Figure \ref{com_dv}, which shows the frequency of the companions to be rising 
towards smaller $\Delta cz$. 
If this trend reflects a physical association between sample SFDG and companion
(see Section \ref{discussion}), then our cutoff, $\Delta cz\,<$\,500\,km\,s$^{-1}$, 
should mainly select true companions, i.e. objects likely to act as perturbers repeatedly 
or on longer timescales, rather than randomly passing interlopers.

Table \ref{biastab} (first row) shows the companion statistics obtained by
the latter search criteria.

We will in the following refer to the galaxies with at least one possible companion within 
the limits stated above as to the {\em non--isolated sample} and to those for which no 
companion candidates were found as to the {\em isolated galaxies}.
This is for the sake of simplicity; one has to keep in mind that the isolation criteria 
described above are empirical, being based on assumptions and observational clues.

%----------------------------------------------------------------------------------
\subsubsection{Companion properties; bias against faint companions}
\label{bias}

The first row of Table \ref{biastab} shows that the majority ($\ga$\,80\,\%) 
of the possible close companions are dwarf galaxies (M$_B >$\,--18\,mag); this is also 
illustrated in Figure \ref{com_b}. Their mean $B-V$ colour (0.44$\pm$0.11\, mag) is nearly 
equal to the average value for the sample galaxies (cf. Table \ref{sf_tab}), pointing to 
recent or ongoing SF activity.
The differences between the B luminosities of each sample galaxy and its companions
are shown in Figure \ref{com_db}. This distribution shows a large scatter around a median 
of 0.72 mag, which suggests that we tend to find primarily the brightest companion galaxies 
(cf. Section \ref{discussion}).

The observational bias against low--luminosity companions, i.e. against the majority of 
close companions of SFDGs, is also evident from Table \ref{biastab}. 
The rough completeness limit of the NED of $\sim$15.5\,B\,mag (cf. Section \ref{data}), 
translates, by the $H_0$ adopted herein (Section \ref{intro}) and assuming a pure 
Hubble flow, to a maximum distance of 50\,Mpc ($cz \approx$\,3760\,km\,s$^{-1}$) 
for a dwarf companion (M$_B >$\,--18\,mag) to be listed in the NED.
Indeed, for $cz \geq $\,4000\,km\,s$^{-1}$, no dwarf companions were found. 
In the determination of the fraction of SFDGs which have a possible dwarf
companion, one is therefore forced to stick to small redshifts. 
This is further demonstrated by the fact that for the subsample at $cz \leq
$\,2000\,km\,s$^{-1}$, the fraction of non--isolated galaxies
is considerably higher (31\,\%) than for the total sample (16\,\%). 
We will further comment on this point in Section \ref{discussion}.

\begin{figure}
\resizebox{\hsize}{!}{\includegraphics{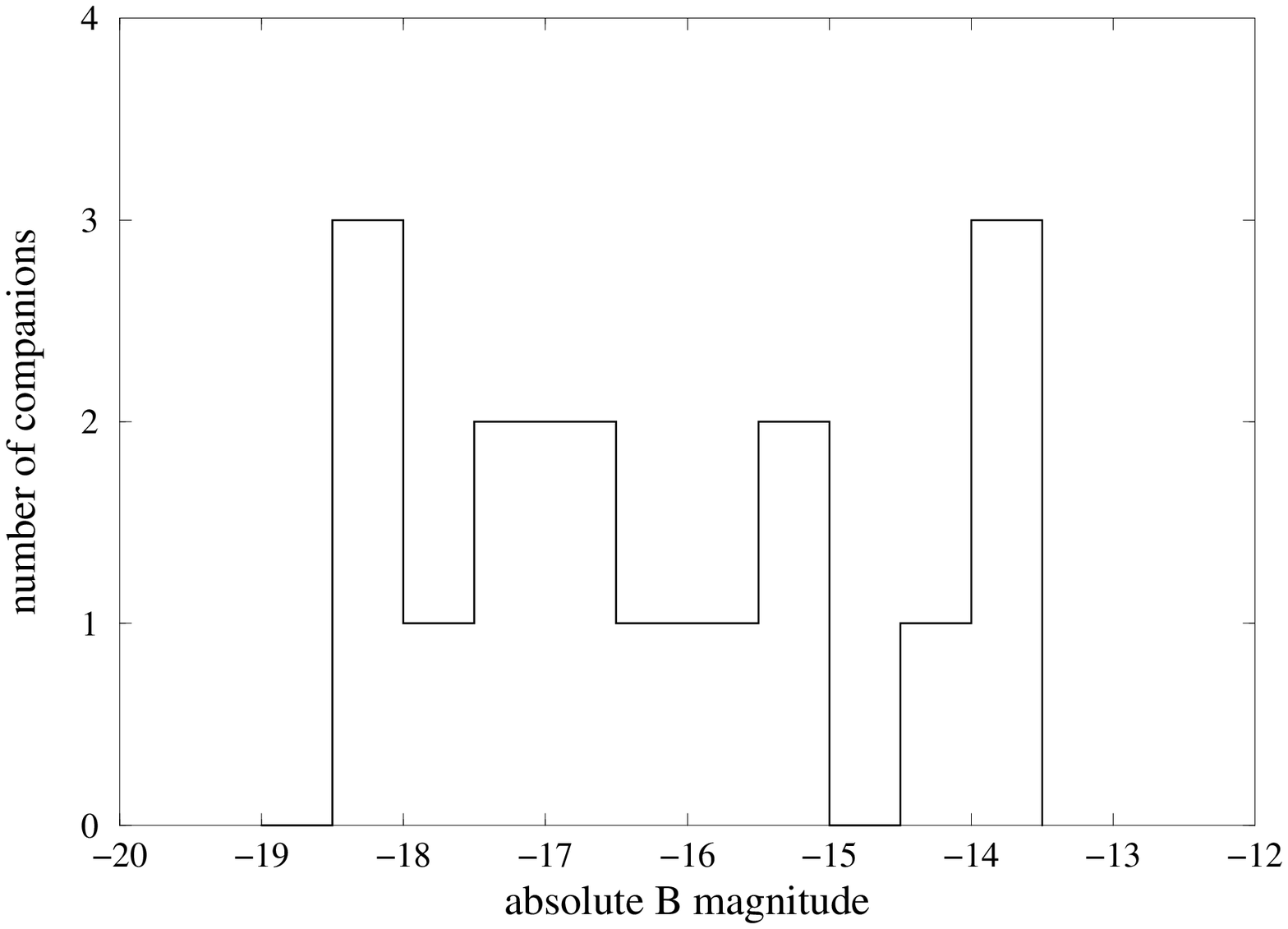}}
\caption{Distribution of the B magnitudes of the SFDG companion candidates.}
\label{com_b}
\end{figure}

\begin{figure}
\resizebox{\hsize}{!}{\includegraphics{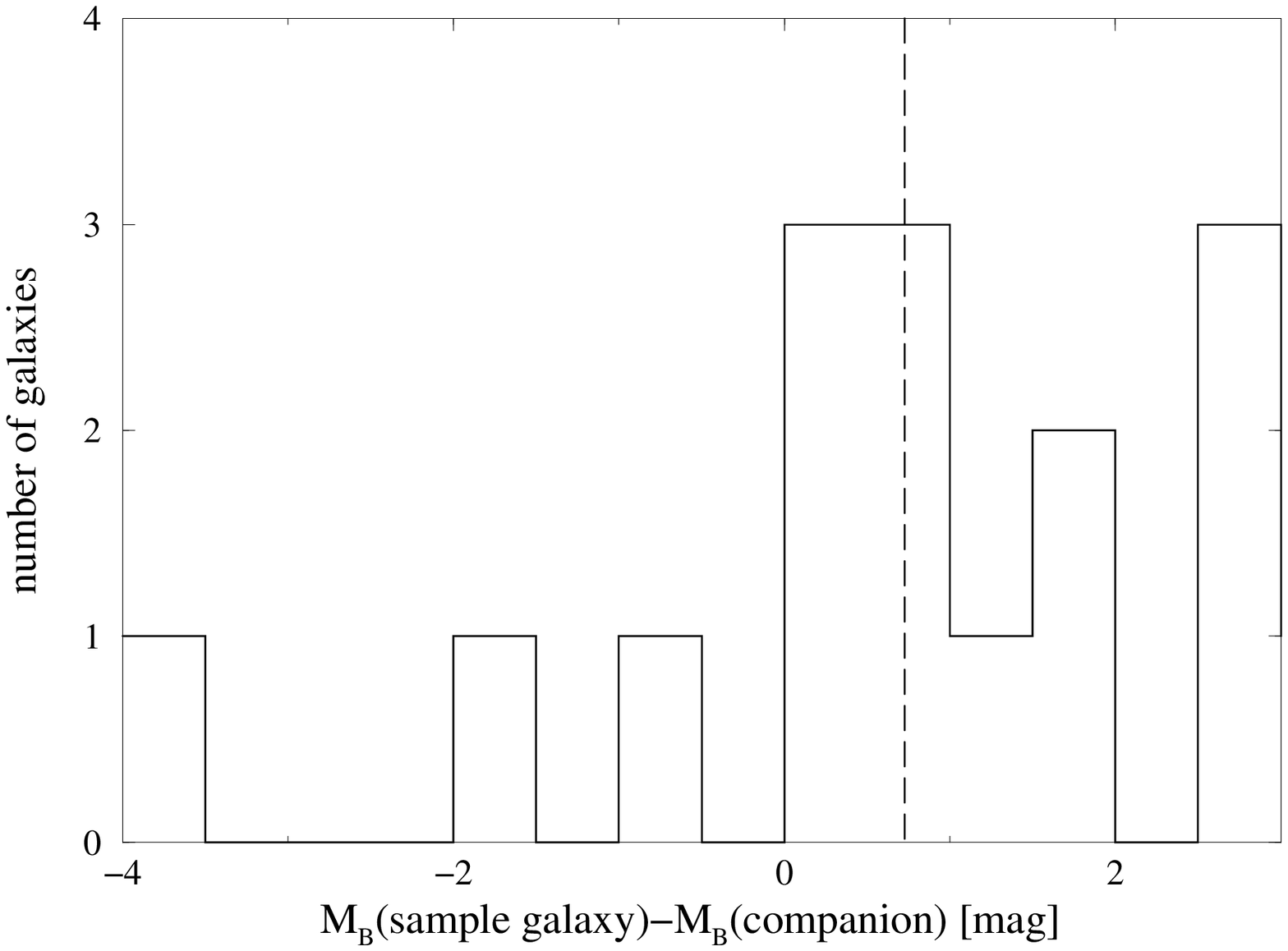}}
\caption{Distribution of differences between the B luminosity of each SFDG and its 
found companion candidate(s). The median value
of the distribution, 0.72 mag, is marked by the dashed line.}
\label{com_db}
\end{figure}

%----------------------------------------------------------------------------------
\subsection{Star Formation and Environment}
%----------------------------------------------------------------------------------

%----------------------------------------------------------------------------------
\begin{table}
\caption{Comparison of the SF activity indicators for different subsamples.}
\label{sf_tab}
\begin{tabular}{lccccc}
\hline\hline
SF indicator                    & isolated & non--isol. & K.--S.\\
(1)                             & (2)      & (3)        & (4)                         \\ \hline
$EW$(H$\beta$)$[{\rm \AA}]$     & 68.4$\pm$15.1 & 94.6$\pm$24.7  &       0.45    \\
\# data points          & 19           & 11             &               \\ \hline
$(B-V)[$mag$]$                  & 0.43$\pm$0.07 & 0.48$\pm$0.06 &       0.90    \\
\# data points          & 27       & 13                 &               \\ \hline\hline 
\end{tabular}
\ \\[1em](2) Mean of the respective SF activity indicator for the
'isolated' subsample with standard deviation about the mean.\\
(3) Like (2), but for the 'non--isolated' subsample.\\
(4) Result of a Kolmogorov--Smirnov test, applied to the distribution of the respective 
SF activity indicator for the 'non--isolated' and 'isolated' subsamples. The probability 
of the hypothesis that both distributions are drawn from the same parent distribution is 
shown; small values suggest that the distributions are different from each other.
\end{table}
%----------------------------------------------------------------------------------
%
\begin{figure*}
\centering
\includegraphics[width=16.8cm]{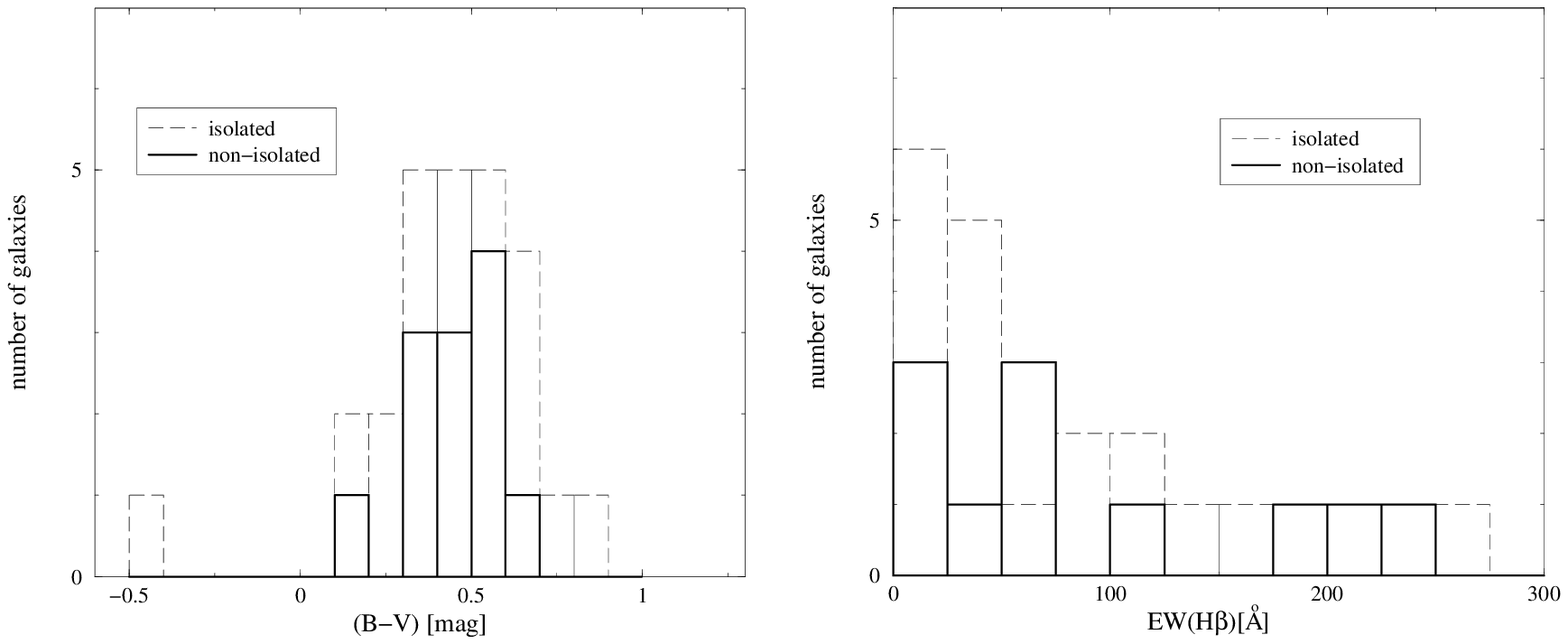}
\caption{Histograms of the {\bf left:} $B - V$ colours and {\bf right:} H$\beta$ equivalent 
widths $EW($H$\beta )$ 
for the isolated (dashed line) and the non--isolated (solid line) subsample of SFDGs. The
comparison is restricted to the sample at close distances, i.e. at redshifts 
$cz\,<$\,2000\,km\,s$^{-1}$.}
\label{histo}
\end{figure*}

To assess whether the SF activity of a SFDG can be influenced by its close environment, we 
chose those commonly used indicators which were available for most of our sample 
galaxies and trace the relative strength of the SF activity on 
different timescales (see e.g. Schaerer \cite{schaerer99} for a review):\\
(i) The $H\beta$ equivalent width $EW($H$\beta )$, approximately relating the flux of 
Lyman continuum photons, i.e. the actual or very recent ($\tau _{\rm burst}\la $ 10 Myr) SFR, 
to the continuum flux in the B band, which is usually dominated by stellar emission.\\
(ii) The integral $B-V$ colour of the galaxies which is an indicator of the
past SF activity on timescales of $\sim$ 10$^8$ yr (see e.g. the 
models by Kr\"uger \cite{krueger92} for different burst parameters).
We refrain from a discussion of IRAS fluxes, originally collected and analyzed
for the sample as well, as they are available for few SFDGs only, and possibly 
biased towards metal--rich dwarfs (see Dultzin--Hacyan et al. \cite{dultzin90}).

In view of the bias against faint companions (Section \ref{bias}) we stick here to 
the SFDGs with redshifts $cz\,<$\,2000\,km\,s$^{-1}$.
The above mentioned SF indicators for the isolated and non--isolated 
samples are compared in Figure \ref{histo}. The corresponding statistics are listed in 
Table \ref{sf_tab}.
Column (4) lists a comparison of the sample distributions 
by means of Kolmogorov--Smirnov tests. 
Both the $EW($H$\beta )$ and $B-V$ colours of the isolated and non--isolated 
subsamples are compatible with the hypothesis that they are drawn from equal parent 
distributions. Similarly, the sample means of $EW($H$\beta )$ and $B-V$ are 
compatible for the isolated and non--isolated subsamples within their respective sample 
standard deviations. 
Hence, from the available data, we do not confirm the higher mean $EW($H$\beta )$ of 
the 'non--isolated' sample.

%---------------------------------------------------------------------- 
\section{Discussion}
\label{discussion}
%-----------------------------------------------------------------------------------
%
\subsection{Frequency, distribution and properties of the companion candidates}
\label{compprops}

The present investigation reveals that a significant fraction of SFDGs possesses 
possible close companions which are in almost all cases dwarf galaxies. 
However, the selection of the nearby SFDG subsample at $cz<$2000\,km\,s$^{-1}$  
does not eliminate, but merely moderates, the degree of incompleteness, 
which is illustrated by the distribution of the B magnitudes of the companion 
candidates (Figure \ref{com_b}): unlike that expected for a complete sample, it
shows no increase towards fainter luminosities. 
In this respect, the median of the magnitude differences between a 
sample galaxy and its found companion candidate(s) ($\sim$\,0.7\,B\,mag, 
Figure \ref{com_db}) should be interpreted as due to selection effects: 
the companions of SFDGs are not necessarily brighter than the SFDGs 
themselves, but only the brightest among the companions are listed in NED.
Nevertheless, even these brightest companions qualify as dwarfs.
The integrated colours of these companions are similar to those of the 
sample SFDGs (cf. Section \ref{bias}), being consistent with a similar 
recent SF history in both sample SFDG and found companions. 
While this could be interpreted as a result of mutual interaction, it might again 
reflect the abovementioned selection effect: faint companions are most 
likely to be detected in emission line surveys (cf. Salzer et al. \cite{salzer89b}), 
which, of course, list preferably actively star--forming galaxies.

The quantification of the true frequency of dwarf companions
suffers from the  principal drawback that
both the low--luminosity end of the galaxy LF and the frequency of \ion{H}{i} companions 
without any optical counterpart (Taylor et al. \cite{taylor94}, 
\cite{taylor95}, \cite{taylor96}) are not yet well--constrained.
Therefore, even a uniform and well--defined homogeneous completeness limit for 
the NED would to date not allow any meaningful extrapolation.
Instead, we emphasize that the value of $\approx$\,30\,\% inferred in Section \ref{bias}  
must be considered a lower limit to the true fraction of SFDGs with close optical 
or purely gaseous (dwarf) companions within $\Delta cz < $\,500\,km\,s$^{-1}$ and 
$s_p\,<$\,0.1\,Mpc.
It must be noted that this work and the one by Pustilnik et al. (\cite{pustilnik00}), 
performed independently and using different approaches and data, find almost equal 
frequencies of low--luminosity companions of SFDGs/BCDs. 
Furthermore, Pustilnik et al. (\cite{pustilnik00}) also report evidence for a 
considerable number of even fainter, but yet uncatalogued companions.
Altogether, this lends further support to the results by \cite{lindner96}, who 
attributed the apparent extreme isolation of some BCDs to an observational bias against 
distant, intrinsically faint companions. 

The small redshift differences between SFDGs and companions, the distribution of which
rises towards smaller values, are comparable to what is found for binary galaxies 
(Schneider \& Salpeter \cite{schneider92}). In particular, 
a similar behaviour is reported for dwarf satellites of field spirals 
(Zaritsky et al. \cite{zaritsky97}).
This is illustrated in Figures \ref{tdv} and \ref{com_dv}, where we plot our results 
of a near--object search (not restricted to dwarf galaxies) in the NED around field 
spirals from Kennicutt \& Kent (\cite{kennicutt83}). Analogous to spiral companions, 
there is no evident correlation between $\Delta cz$ and $s_p$ for the SFDG companions.
A plausible explanation of this issue was offered in the framework of a CDM scenario,  
taking into account a number of further dynamical arguments, by Zaritsky et al. 
(\cite{zaritsky97}). They consider the spiral satellites to be associated with an extended 
massive DM halo of the mother galaxy, which strongly influences the system's dynamics, so that 
$\Delta cz$ is widely independent of $s_p$. 
Figure \ref{tdv} suggests that $\Delta cz$ and $s_p$ are also uncorrelated for the 
SFDG--companion systems. If this can be corroborated by a larger dataset, then a hypothesis 
 worth investigating for compatibility with CMD models is that SFDGs and their companions 
 share common DM haloes of similar mass and structure.

At least, the data strongly suggest some physical association between SFDGs and
their companions. In combination with our considerations on the frequency of dwarf 
companions, this leads to the assumption that SFDGs are frequently 
{\em associated} with systems of two or more dwarf galaxies. This is reminiscent of 
the dwarf galaxy groups \cite{lindner96} found in voids, and may offer important 
clues to dwarf galaxy formation.

\subsection{Close dwarf companions as starburst triggers?}

The similarity of the SF tracers for the 
isolated and non--isolated subsamples may suggest that the observed 
starburst activity is regulated by solely internal processes. This hypothesis can, on the 
other hand, not be proven, and neither can an influence of the presence or absence of 
companions be disproven. Indeed, this latter hypothesis may receive some support from 
the marginal trend in $EW$(H$\beta$) that the data shows.
The large scatter of the analyzed SF diagnostics is not surprising, for a number of reasons:

(i) The severe incompleteness of the available data aggravates the separation of
galaxies with and without close companions.

(ii) Balmer emission lines, i.e. Lyman continuum tracers, correspond to short--term 
variations in the SFR. The latter are particularly strong in SFDGs, where SF generally occurs 
in short bursts of $\la$\,10$^7$ yr (Elmegreen et al. \cite{elmegreen96}). 
Considerable variations with time also occur for the $B-V$ colours after the onset of
a starburst (see e.g. Kr\"uger \cite{krueger92}). Whatever triggers a starburst, the time 
window during which strongly enhanced SF indicators are observable is narrow.
In addition, such SFR tracers also depend on evolutionary parameters -- metallicity, 
age and stellar continuum contribution -- of the SFDG itself.
 
The hypothesis of interaction--induced starburst activity in SFDGs might be more 
reliably assessed when a large sample of SFDGs with sufficiently deeply observed 
environments is available. 
However, the abovementioned intrinsic sources of scatter certainly limit the usefulness of 
a comparison that is restricted to integral SFR tracers. 

This research emphasizes the need for a systematic effort on the part of theory
to advance the understanding of conditions required for inducing and 
sustaining starburst activity in dwarf galaxies, with particular emphasis on dwarf
galaxy interactions.

%---------------------------------------------------------------------------------
\section{Summary and Conclusions}
\label{summary}
%----------------------------------------------------------------------------------

We have searched the close environment of a sample of 98 star--forming dwarf 
galaxies (SFDGs) in field and low--density environments, taken from different catalogues, 
for possible companion galaxies. 
To supplement previous work, which had mostly dealt with luminous companions, 
a subsample of nearby SFDGs was chosen to moderate the effect of observational bias 
against low luminosity companions. Distances were obtained from the recession velocity 
of the sample galaxies assuming a pure Hubble flow and have been corrected
for Virgocentric infall.
Using the Nasa Extragalactic Database (NED) as a search catalogue, objects 
with a  projected linear separation of $<$100 kpc and a recession velocity difference 
$<$2000\,km\,s$^{-1}$  were catalogued as possible companions of a SFDG.
Most of them differ in their recession velocity by
$<$500\,km\,s$^{-1}$ from the sample SFDGs, which, along with a number of 
other considerations, led us to adopt this value as a limit to identify possible companions. 
For both SFDGs and companion candidates, spectrophotometric data were 
compiled from the NED and a number of literature sources.

We studied the frequency, redshift difference distribution, and photometric 
properties of companion candidates, as well as their possible influence 
on the star formation (SF) activity of the sample SFDGs.
Our results can be summarized as follows:

(i) A substantial fraction of SFDGs possess companion galaxies within its 
close environment. The overwhelming majority of these companions are dwarfs
(M$_{\rm B}>$\,--18), which renders their detection at large distances difficult.
The fraction of SFDGs for which we detect dwarf companion candidates,
$\approx$\,30\,\%, must be considered a lower limit to  the true value, given the 
increasing incompleteness of the available data set for systems fainter than 
$m_B\sim$\,15.5\, mag. A meaningful extrapolation to fainter magnitudes is 
precluded by the poorly constrained frequency of \ion{H}{i} companions with
no optical counterpart, and by the uncertain faint end of the galaxy luminosity function.

(ii) The recession velocity differences between SFDGs and identified companions amount 
typically to $\la$250\,km\,s$^{-1}$. Their frequency rises towards lower differences. 
This is similarly reported for dwarf companions of spirals, and suggests that the SFDG 
companions are physically associated with the sample galaxies.

(iii) Both the $B-V$ colours and the H$\beta$ equivalent 
widths appear compatible for objects with and without a possible perturber. 
The significance level of the data is too low to prove the
hypothesis of a solely internal regulation of the SF activity, whereas  
 external influence cannot be disproven either.
The considerable intrinsic scatter inherent to statistical studies based 
on the abovementioned SF diagnostics, together with the insufficiently 
complete companion search databases, blurs the picture.

The identification of companions in the close environment of SFDGs 
readdresses the question of whether gravitational interactions are 
partly responsible for the onset of SF activity in gas-rich dwarfs.
Whereas interaction--induced SF activity was dismissed for most SFDGs due to 
the absence of luminous companions, the presence of faint dwarfs 
in the close vicinity of SFDGs provides again support to this hypothesis. 
While deep radio and spectrophotometric surveys 
may yield more conclusive answers, it will equally be important to  
theoretically model the effects of interactions on dwarf galaxies.

%----------------------------------------------------------------------------------
\begin{acknowledgements}

KGN gratefully acknowledges financial support from the German Research Foundation 
(DFG) grant FR325/50-1, and from the IAC Summer Research Programme 1998.
This study was partly financed by the Spanish DGES (Direcci\'{o}n General de 
Ense\~{n}anza Superior) (grant PB97-0158). PP received support from Deutsche 
Agentur f\"ur Raumfahrtangelegenheiten (DARA) GmbH grant 50 OR 9907 7. 
This Research has made use of the NASA/IPAC Extragalactic Database (NED)
which is operated by the Jet Propulsion Laboratory, CALTECH, under
contract with the National Aeronautic and Space Administration.
We have made use of the Lyon-Meudon Extragalactic Database (LEDA)
supplied by the LEDA team at the CRAL-Observatoire de Lyon (France).
We thank S.A. Pustilnik and A. Kniazev for fruitful discussions and
helpful comments on this paper. KGN wishes to thank L.M. Cair\'os, J. Iglesias, 
J. V\'{\i}lchez and the IAC staff for their hospitality. We thank the
referee, Dr. Metcalfe, for his helpful comments. 

\end{acknowledgements}

%-----------------------------------------------------------------------
\begin{table*}
\centering
\caption[]{Sample list: (1) name; (2) recession velocity in km\,s$^{-1}$, corrected for peculiar velocities introduced by Virgo Cluster perturbations (see Section \ref{vc_correction}); (3) integral $B - V$ colour; (4) H$\beta $ equivalent width (in emission); (5) number of close companion candidates found
in the NED within the limits described in Section \ref{distribution}. }
\label{samtab}
\begin{tabular}{lrrrrclrrrr}
\cline{1-5}\cline{7-11}
Name & $cz^{corr.}$ & $B-V$ & $EW($H$\beta)$& \# Comp. & \ \ \ \ \ \ \ \ \ \ & Name & $cz^{corr.}$ & $B-V$ & $EW($H$\beta)$& \# Comp.\\
 & km s$^{-1}$ & mag & ${\rm \AA}$ & & & & km s$^{-1}$ & mag & ${\rm \AA}$ & \\ 
(1) & (2) & (3) & (4) & (5) & & (1) & (2) & (3) & (4) & (5) \\ 
\cline{1-5}\cline{7-11}
UM 306&  4985&  0.42&  26.4& 0 & & UM 523&  1620&  0.51&  16.9& 1 \\ 
UM 92&   6949&  0.33&  38.3& 0 & & UM 533&   737&  0.35& 206.5& 1 \\ 
UM 323&  1998&  0.38&  25.9& 0 & & UM 538&   855&  0.56&  41.2& 0 \\          
UM 330&  5141&  0.57&  18.8& 0 & & UM 539&  6065&  0.60&  56.5& 0 \\          
UM 334&  4924&  0.32&  11.4& 0 & & UM 549&  5642&  0.36&  15.5& 0 \\          
UM 335&  4984&  0.13&   ---& 0 & & UM 552&  7190&  0.48&  20.9& 0 \\          
UM 336&  5816&  0.24&  54.9& 0 & & UM 559&  1037&  0.18& 261.7& 0 \\          
UM 345&  5754&  0.57&  25.7& 0 & & UM 562&  5290&  0.91& 199.5& 0 \\          
UM 351&  7487&  0.55&  44.5& 0 & & UM 564& 13668&  0.86& 101.1& 0 \\          
UM 369&  5812&  0.40&  58.2& 0 & & UM 570&  6543&  1.18& 199.2& 0 \\          
UM 371&  5536&  0.49&   6.5& 0 & & UM 588&  3581&  0.47&  18.6& 1 \\          
UM 372&  1752&  0.31&  92.7& 0 & & UM 591& 15642&  0.61&  38.6& 0 \\          
UM 374&  5735&  0.59&  34.6& 0 & & UM 597&  6441&  0.46&  22.7& 0 \\          
UM 379&  8227&  0.78&  26.6& 0 & & UM 605&  4956&  0.59&   6.1& 0 \\          
UM 151&  4890&  0.46&   9.8& 0 & & UM 612&  4414&  0.54&  19.1& 0 \\          
UM 382&  3786&  0.39& 124.2& 0 & & UM 618&  4188&  0.39&  39.5& 0 \\          
UM 396&  6208&  1.00& 123.5& 0 & & UM 619&  4455&  0.63&  10.4& 0 \\          
UM 404&  3712&  0.45&  74.1& 1 & & UM 626&  3335&  0.57&   5.1& 0 \\          
UM 406& 11251&  0.92& 104.1& 0 & & UM 628&  7052&  0.15&  32.3& 0 \\          
UM 408&  3637&  0.53&  46.8& 0 & & UM 635&  7192&  0.47&  25.9& 0 \\          
UM 410N& 6960&  0.62&  35.4& 1 & & UM 648&  9715&  0.78&   4.4& 0 \\          
UM 411& 11748&  0.78& 480.5& 0 & & UM 649&  7554&  0.58&  81.7& 0 \\          
UM 417&  2846&  0.19&  63.6& 0 & & UGC 47&   873&  0.34&   ---& 1 \\          
UM 422&  2017&  0.21& 328.6& 0 & & TOL 0127-397&  4797&  0.59&  38.0& 0 \\    
UM 439&   949&  0.27&  41.2& 0 & & DDO 19&   735&  1.21&   ---& 0 \\          
UM 442&  7707&  0.41&  24.8& 0 & & MKN 370&   840&  0.63&  10.0& 0 \\         
UM 444&  6332&  0.65&  36.2& 0 & & MKN 600&   945&  0.46&   ---& 0 \\         
UM 446&  1418&  0.45&  34.8& 0 & & UGC 2432&   764&  0.89&   ---& 0 \\        
UM 452&  1219&  0.63&  13.5& 0 & & UGC 2482&  2646&  0.61&   ---& 0 \\        
UM 454&  3592&  0.42&  21.4& 0 & & UGC 2809&  1239&  0.49&   ---& 0 \\        
UM 455&  3657&  0.36&  59.7& 0 & & c0341-4045&  4497&  0.46&   ---& 0 \\      
UM 456&  1518&  0.34&  42.7& 0 & & DDO 34&   615& -0.40&   ---& 0 \\          
UM 461&   735&  0.43& 241.2& 1 & & MKN 5&   792&  0.50& 150.0& 0 \\           
UM 462&   855&  0.44&  72.6& 1 & & UGC 3516&  1287&  0.65&   ---& 0 \\        
UM 463&  1131&  0.45&  98.5& 0 & & UGC 3658&  1267&  0.29&   ---& 0 \\        
UM 465&  1155&  0.59&  15.7& 1 & & MKN 86&   675&  0.66&   6.2& 2 \\                  
UM 471& 10185&  0.76&  41.8& 0 & & I Zw 18&  1072&  0.19&  69.0& 0 \\               
UM 483&  2099&  0.92&  27.7& 0 & & MKN 36&   517&  0.46&  70.0& 1 \\          
UM 487& 14419&  0.50&   6.6& 0 & & VII ZW 403&   105&  ---&   ---& 0 \\       
UM 490&  5376& -0.15& 113.0& 0 & & NGC 4670&   825&  0.39&   ---& 0 \\        
UM 491&  1753&  0.42&  14.6& 0 & & MKN 209&   352&  0.53&  60.3& 2 \\         
UM 495&  7477&  0.52&  47.7& 0 & & II Zw 70&  1732&  0.52&  49.0& 1 \\        
UM 496& 11479&  0.73& 110.9& 0 & & II Zw 71&  1785&  1.23&   ---& 1 \\        
UM 500&  1805&  0.15& 117.8& 1 & & PHL 293 B&  1762&  0.77& 110.0& 0 \\       
UM 501&  1745&  0.39& 184.5& 1 & & MKN 324&  1635&  0.60&   3.8& 0 \\           
UM 504&  1867&  0.50&  17.9& 0 & & MKN 67&  1680&  0.31& 105.0& 0 \\          
UM 507&  6094&  0.58&  94.2& 0 & & MKN 169&  1777&  0.51&  21.9& 0 \\    
UM 512&  4488&  0.51&  13.8& 0 & & I Zw 123&  1125&  0.60& 144.5& 0 \\        
UM 513&  3457&  0.82&   7.3& 0 & & HARO 38&  1327& ---&   ---& 0 \\

\cline{1-5}\cline{7-11}
\end{tabular}
\end{table*}
%-----------------------------------------------------------------------

%----------------------------------------------------------------------------
\end{document}